\documentclass[12pt,english]{article}
\usepackage{RR}
\RRNo{6372}
\usepackage[T1]{fontenc}
\usepackage[latin1]{inputenc}
\usepackage{amsmath}
\usepackage{graphicx}
\usepackage{setspace}
\newcommand{\real}{I\!\!R}

\usepackage{babel}
\RRdate{Novembre 2007}
\RRauthor{M.H. Tber\thanks[inriasop]{INRIA Sophia Antipolis, France} \and L. Hasco\"et\thanksref{inriasop} \and A. Vidard\thanks[inriagre]{INRIA Grenoble, France} \and B. Dauvergne\thanksref{inriasop}}
\authorhead{Tber et al.}
\RRetitle{Building the Tangent and Adjoint codes of the Ocean
General Circulation Model OPA with the Automatic Differentiation tool TAPENADE}
\RRtitle{Construction des codes Tangent et Adjoint du mod\`ele de circulation g\'en\'erale oc\'eanique OPA par l'outil de Diff\'erentiation Automatique TAPENADE}
\titlehead{Tangent and Adjoint differentiation of OPA}
\RRresume{Le mod\`ele de circulation g\'en\'erale oc\'eanique OPA
est d\'evelopp\'e par l'\'equipe LODYC de l'universit\'e Paris VI.
La nouvelle version 9 d'OPA constitue une \'evolution majeure, avec en particulier une
migration vers FORTRAN95. Les codes Lin\'eaire Tangent et Adjoint d'OPA,
qui auparavant \'etaient \'ecrits \`a la main, doivent donc \^etre
red\'evelopp\'es. Nous utilisons l'outil de Diff\'erentiation Automatique TAPENADE
pour construire les codes Tangent et Adjoint d'OPA 9.
Nous validons les d\'eriv\'ees obtenues par comparaison avec les
Diff\'erences Divis\'ees et sur deux applications test incluant des exp\'eriences jumelles.
Nous utilisons le sch\'ema de checkpointing r\'ecursif binomial de Griewank et Walther
pour am\'eliorer les performances du code adjoint.
Nous utilisons une strat\'egie sp\'ecifique pour diff\'erentier le
solveur lin\'eaire it\'eratif provenant du sch\'ema implicite d'avancement en temps.
Nos r\'esultats montrent un co\^ut raisonnable, tant en consommation m\'emoire que 
pour le temps d'ex\'ecution de l'adjoint.}
\RRabstract{The ocean general circulation model OPA is developed by the LODYC team
at Paris VI university. OPA has recently undergone a major
rewriting, migrating to FORTRAN95, and its adjoint code needs to be rebuilt.
For earlier versions, the adjoint of OPA was written by hand at a high development cost.
We use the Automatic Differentiation tool TAPENADE to build
mechanicaly the tangent and adjoint codes of OPA. We validate
the differentiated codes by comparison with divided differences,
and also with an identical twin experiment.
We apply state-of-the-art methods to improve the performance
of the adjoint code. In particular we implement the Griewank and
Walther's binomial checkpointing algorithm which gives us an optimal trade-off
between time and memory consumption. We apply a specific strategy
to differentiate the iterative linear solver that comes from the implicit
time stepping scheme.}
\RRmotcle{OPA, Circulation Oc\'eanique, TAPENADE, Diff\'erentiation Automatique, mode inverse, Code Adjoint, Checkpointing}
\RRkeyword{OPA, general circulation model, TAPENADE, Automatic Differentiation, reverse mode, Adjoint Code, Checkpointing}
\RRprojet{Tropics}
\RRtheme{\THNum}
\URSophia

\begin{document}
\makeRR

\section{Introduction}

The development of tangent and adjoint models is
an important step in addressing sensitivity analysis and variational
data assimilation problems in Oce\-anography.
Sensitivity analysis is the study of how model output
varies with changes in model inputs. The sensitivity information
given by the adjoint model is used directly to gain an understanding
of the physical processes. In data assimilation, one considers a cost
function which is a measure of the model-data misfit. The adjoint
sensitivities are used to build the gradient for descent algorithms.
Similarly the tangent model is used in the context
of the incremental algorithms \cite{CourtierEtAl} to linearize the
cost function around a background control. For the previous version
8 of the Ocean General Circulation Model  OPA \cite{MadecEtAl}, Weaver
et al \cite{WeaverEtAl} developed the numerical tangent and
adjoint codes by hand using classical techniques \cite{GeiringKaminski,Talagrand}.
Since then, the OPA model has undergone a major update. Particularly
the new versions are fully rewritten in FORTRAN95. In this paper,
we report on the development of tangent and adjoint codes
of OPA using the Automatic Differentiation (AD)
tool TAPENADE \cite{Hascoet-Pascual}. A brief description of the
OPA model and the configuration used in this work is given in the
next section. In section 3 we present the principles of AD and how they are reflected into the functionalities
of the AD tool TAPENADE.
In section 4 we focus on the most interesting difficulties that we encountered  in the application
of AD to such a large code. Section 5 shows some experiments
that validate our derivatives and presents two illustrative applications,
focusing on computational aspects rather than implications
for oceanography. An outlook of further work is given in
the conclusion.

\section{The Ocean General Circulation Model OPA}

Developed by the LODYC team at Paris VI university,
OPA is a flexible ocean circulation model that
can be used either in a regional or in a global ocean configuration.
OPA is the ocean model component of NEMO
(Nucleus For European Modelling of the Ocean) and is
widely used in the scientific community.
Moreover it is becoming a major actor in operational
oceanography (Mercator, ECMWF, UK-Met office)
Its formulation is based on the so-called primitive equations for
the temporal evolution of ocean velocity currents,
temperature and salinity in
its three horizontal and vertical dimensions. These equations are
derived from Navier-Stokes equations coupled with a state equation
for water density and heat equation, under Boussinesq and hydrostatic
approximations. 

Let us introduce the following variables: $\mathbf{U}$ the velocity
vector, $\mathbf{U}=\mathbf{U}_{h}+w\mathbf{k}$ (the subscript
$h$ denotes the local horizontal vector), $T$ the potential temperature,
$S$ the salinity, $p$ the pressure and $\rho$ the in-situ density.
The vector invariant form of the primitive equations in an orthogonal
set of unit vectors linked to the earth are written as follows
\[\left\{\begin{array}{lcl}
\dfrac{\partial\mathbf{U}_{h}}{\partial t\vphantom{P_P}} & = & -\left[\left(\nabla\times\mathbf{U}\right)\times\mathbf{U}+\dfrac{1}{2}\nabla\left(\mathbf{U}\right){}^{2}\right]_{h}-f\mathbf{k}\times\mathbf{U}_{h}-\dfrac{1}{\rho_{0}}\nabla_{h}p+\mathbf{D}^{\mathbf{U}}\\
\dfrac{\partial p\vphantom{P^P}}{\partial z\vphantom{P_P}} & = & -\rho g\\
\nabla\cdot\mathbf{U}\vphantom{P^P P_P} & = & 0\\
\dfrac{\partial T\vphantom{P^P}}{\partial t\vphantom{P_P}} & = & -\nabla\cdot\left(T\mathbf{U}\right)+D^{T}\\
\dfrac{\partial S\vphantom{P^P}}{\partial t\vphantom{P_P}} & = & -\nabla\cdot\left(S\mathbf{U}\right)+D^{S}\\
\rho\vphantom{P^P} & = & \rho\left(T,S,p\right)\end{array}\right.\]
 where $\nabla$ is the generalized derivative vector operator, $t$
the time, $z$ the vertical coordinate, $\rho_{0}$ a reference density,
$f$ the Coriolis acceleration, and $g$ the gravity acceleration.
$\mathbf{D}^{\mathbf{U}},$ $D^{T}$ and $D^{S}$ are the parametrization
of small scale physics for momentum, temperature and salinity, including
surface forcing terms. 
A full description
of the model basics, discretization, physical and numerical details can
be found in \cite{MadecEtAl}. 

Through this paper, OPA is used in its global free surface configuration
ORCA-2. In this configuration the model uses a rotated grid with poles
on North America and Asia in order to avoid the singularity problem
on the North Pole. The space resolution is roughly equivalent to a
geographical mesh of 2° by 1.3° with a meridional resolution of 0.5°
near the Equator (see figure \ref{ORCA-2}). The Vertical domain,
spreading from the surface to a depth of 5000m, is meshed using 31
levels with levels 1 to 10 in the top 100 meters. The time step is
96 minutes so that there are 15 time steps per day. The model is
forced by heat, freshwater, and momentum fluxes from the atmosphere
and/or the sea-ice. The solar radiation penetrates the upper layers
of the ocean. Zero fluxes of heat and salt are applied through the
bottom. On the lateral solid boundaries a no-slip condition is also
applied. Initialization of the model for temperature and salinity
is based on the Levitus et al. (1998) climatology with a null initial
velocity field. For more details about the space time-domain and the
ocean physics of ORCA-2, we refer to the page dedicated to this configuration
in the official website of NEMO-OPA%
\footnote{http://www.lodyc.jussieu.fr/NEMO/general/description/ORCA\_config.html%
}.

The configuration ORCA-2 is routinely used by MERCATOR/Meteo-France
to compute the oceanic component of their seasonal forecasting system.
The size of OPA-9, 200 modules defining 800 procedures with
over 100 000 lines of FORTRAN95, makes it the largest
application differentiated by TAPENADE to date.
The computational kernel which is actually differentiated
accounts for 330 procedures.

\begin{center}%
\begin{figure}
\begin{centering}\includegraphics[width=14cm,height=8cm,keepaspectratio]{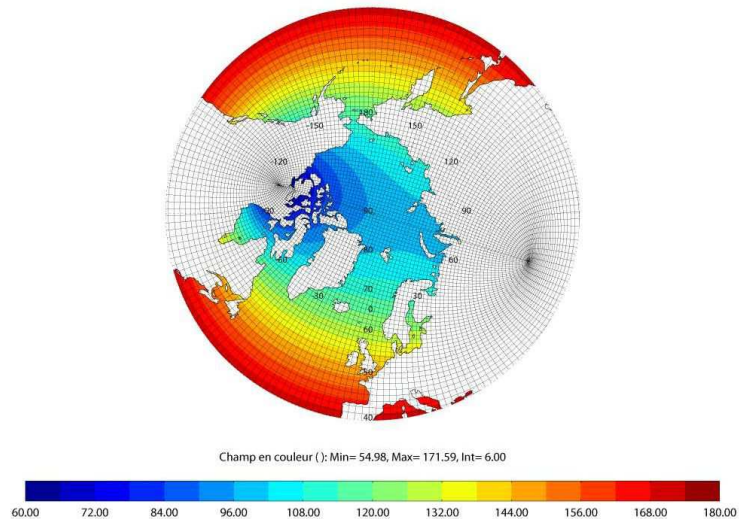}\par\end{centering}

\caption{\label{ORCA-2}ORCA 2 Mesh}
\end{figure}
\par\end{center}

\section{Principles of AD and the tool TAPENADE}

TAPENADE \cite{Hascoet-Pascual} is an AD tool
developed by the Tropics\footnote{
\tt http://www-sop.inria.fr/tropics/}
team at INRIA.
Given the source of an original program
that evaluates a mathematical function,
and given a selection
of input and output variables to be differentiated,
TAPENADE produces a new source program that computes the
partial derivatives of the selected outputs with respect
to the selected inputs.

Basically, TAPENADE does that by inserting
additional statements into a copy of the original program.
Like other AD tools, TAPENADE is based on the
fundamental observation that the original program {\tt P},
whatever its size and run time, computes a function
$F, X\!\!\in\!\!\real^{m} \mapsto Y\!\!\in\!\!\real^{n}$
which is the composition of the elementary functions
computed by each run-time instruction. In other words if
{\tt P} executes a sequence of elementary statements
$I_k, k\in[1..p]$, then {\tt P} actually evaluates
$$F = f_p \circ f_{p-1} \circ \dots \circ f_1 \enspace,$$
where each $f_k$ is the function implemented by $I_k$.
Therefore one can apply the chain rule of derivative
calculus
to get the Jacobian matrix $F'$, i.e. the partial
derivatives of each component of $Y$ with respect to
each component of $X$. Calling $X_0=X$ and
$X_k=f_k(X_{k-1})$ the successive values of all
intermediate variables, i.e. the successive {\em states} of
the memory throughout execution of {\tt P}, we get
\begin{equation}\label{eqchainrule}
F'(X) = f'_p(X_{p-1}) \times f'_{p-1}(X_{p-2}) \times
\dots \times f'_1(X_0) \enspace.
\end{equation}
The derivatives $f'_k$
of each elementary instruction are easily built, and must
be inserted in the differentiated program so
that each of them has the values $X_{k-1}$ directly available
for use.
This process yields analytic derivatives,
that are exact up to numerical accuracy.

In practice, two sorts of derivatives are of
particular importance in scientific computing: the
tangent (or directional) derivatives, and the
adjoint (or reverse) derivatives.
In particular, tangent and adjoint
are the two sorts of derivative programs required
for OPA, and TAPENADE provides both.
The tangent derivative is the product
$\dot{Y} = F'(X) \times \dot{X}$ of the full Jacobian times
a direction $\dot{X}$ in the input space.
From equation (\ref{eqchainrule}), we find
\begin{equation}\label{eqtgtmode}
\dot{Y} = F'(X) \times \dot{X}
= f'_p(X_{p-1}) \times f'_{p-1}(X_{p-2}) \times
\dots \times f'_1(X_0) \times \dot{X}
\end{equation}
which is most cheaply executed from right to left
because matrix$\times$vector products are much cheaper
than matrix$\times$matrix products.
This is also the most convenient execution order because
it uses the intermediate values $X_k$ in the same order
as the program {\tt P} builds them.
On the other hand the adjoint derivative is the product
$\overline{X} = F'^{*}(X) \times \overline{Y}$ of
the {\em transposed} Jacobian times a weight vector
$\overline{Y}$ in the output space. The
resulting $\overline{X}$ is the gradient of the
dot product $(Y \cdot \overline{Y})$.
From equation (\ref{eqchainrule}), we find
\begin{equation}\label{eqadjmode}
\overline{X} = F'^{*}(X) \times \overline{Y} =
f'^{*}_1(X_0) \times \dots \times f'^{*}_{p-1}(X_{p-2})
\times f'^{*}_p(X_{p-1}) \times \overline{Y}
\end{equation}
which is also most cheaply executed from right to left.
However, this uses the intermediate values $X_k$ in the
inverse of their building order in {\tt P}.

Regarding the runtime cost for obtaining the derivatives,
both tangent $\dot{Y}$ and adjoint $\overline{X}$
cost only a small multiple of the original program {\tt P}.
The slowdown factor is less than 4 in theory.
In practice it can be less than 2 for the tangent,
whereas it can reach up to 10 for the adjoint
for a reason discussed below. Despite its higher cost,
the adjoint code is still by large
the cheapest way to obtain gradients.
To get the gradient with the tangent mode would require
$m$ runs of the tangent code, one per dimension of $X$,
whereas this cost is independent from $m$ with
the adjoint mode.

The difficulty of the adjoint mode lies in the fact
that it needs the intermediate values $X_k$ in reverse
order. To this end, TAPENADE basically uses a two-sweeps strategy,
called ``Store-All''. In the first sweep
(the ``forward sweep''), a copy of the original program
{\tt P} is run, together with ``Push'' statements that
store intermediate values on a stack just before they get
overwritten. In the second sweep (the ``backward sweep''),
the derivative statements compute the elementary
derivatives $f'^{*}_{k}(X_{k-1})$ for $k=p$ down to $1$,
using ``Pop'' statements to restore the intermediate
values as they are required. This incurs a cost in
memory space as the maximum stack size needed is attained
at the end of the forward sweep, and is thus
proportional to the length of the program {\tt P}.
There is also a runtime penalty for these
stack manipulations.
TAPENADE implements a number of
strategies~\cite{tapenadeDataFlow} to mitigate
this cost, based on static data-flow analysis of
the program's control flow graph,
reducing the number of values $X_k$ that need to be stored.
However for very long programs such as OPA,
involving unsteady simulations,
Store-All can not work alone.
TAPENADE combines it with a storage/recomputation
trade-off called ``checkpointing''.

\begin{figure}[htbp]
\begin{center}
  \includegraphics[width=\linewidth]{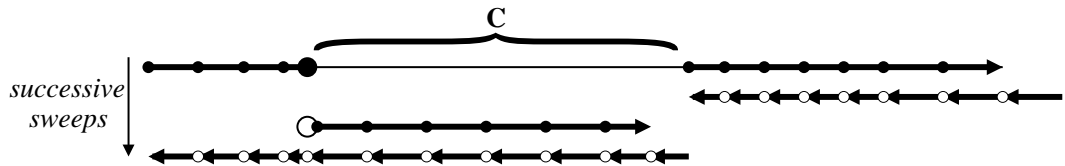}
  \caption{Checkpointing applied to the program piece $C$.
  Rightwards arrows represent forward sweeps, thick when
  they store intermediate values on the stack, thin
  otherwise. Leftwards arrows represent backward sweeps.
  Black dots are stores, white dots are retrieves.
  Small dots are Push and Pops, big dots are snapshots.
  }
  \label{figCkp}
\end{center}
\end{figure}

Checkpointing reduces the maximum
stack size at the cost of duplicated executions.
Consider a piece $C$ of the original program {\tt P}.
Checkpointing $C$ as illustrated on figure~\ref{figCkp}
means that during the main forward sweep,
$C$ pushes no value on the stack. When the backward
sweep reaches back to the place where intermediate
values are now missing on the stack, it runs $C$ a second
time, this time with the ``Store-All'' strategy i.e.
pushing values on the stack.
The backward sweep can then resume safely.
To run $C$ twice requires that enough of its input values,
a ``snapshot'', are stored but the size of a snapshot
is generally much less than the stack size used by $C$.
Obviously, this also slows down the adjoint program.
When $C$ is well chosen, checkpointing can divide
the peak size of the stack by a factor of two.
Checkpoints can be nested, in which case both the stack's
peak size and the adjoint runtime slowdown can grow
as little as the logarithm of the size of {\tt P}.
In its default mode, TAPENADE applies checkpointing to
each procedure call.

TAPENADE capacity to generate robust and efficient tangent
and adjoint codes has been demonstrated on several
real-world test applications~\cite{KimEtAl2,Giles,CastaingsEtAl,SonicBoom,LauvernetEtAl}.
Regarding the application language, it can handle
programs written in FORTRAN.
Taking into account the new programming constructs
provided by FORTRAN95 has required an important
programming effort in the past few years,
mostly to handle modules, structured data types,
array notation, pointers, and
dynamic memory allocation.
Since the new OPA 9 is now written in
FORTRAN95, differentiation of OPA is a
very realistic test for the new TAPENADE 2.2.

There exist several other AD tools.
Restricting to the tools which, like TAPENADE,
operate by source transformation,
provide tangent and adjoint modes, use global
program analysis to optimize the differentiated code,
and have demonstrated their applicability on large
industrial codes, we can mention TAF~\cite{refTAF}
a pioneer of AD for meteorology, now the standard AD tool
for the popular MIT Global Circulation Model.
Unlike TAPENADE's, the adjoint mode of TAF
regenerates the intermediate values $X_k$ by
recomputation from an given initial point.
This is called a ``Recompute-All'' strategy.
Comparison with Store-All strategy is getting blurred by
nested checkpointing, as the adjoint codes grow
more alike as more checkpoints are inserted.
OpenAD~\cite{refOpenAD}, successor of ADIFOR
and ADIC, uses the Store-All strategy.
There are experiments to also apply OpenAD to the MIT GCM.
The tool Adol-C~\cite{refAdolc}, although using
operator overloading instead of source transformation,
is very popular and has been applied successfully to many
industrial applications. Its adjoint mode
can be seen as an extension of the Store-All strategy:
not only the intermediate values are stored on the stack,
but also the computation graph to be differentiated.
This allows the AD tool to perform further optimizations
on this graph, at the cost of a higher memory consumption.

\section{Applying TAPENADE to OPA}

We generated working tangent and adjoint codes for the computational kernel of OPA, using TAPENADE.
Depending on the final application ({\em cf} section 5),
the actual function to differentiate as well as the input and output variables may be different,
but the technical difficulties that we encountered are essentially the same.
This section describes these points. 

\subsection{FORTRAN95 constructs}

The new OPA 9 uses extensively the modular constructs of FORTRAN95.
We had to extend the call-graph internal representation of TAPENADE
to handle the nesting of modules and procedures. Essentially
this nesting is mirrored into the differentiated code.

Because a module can define private components, subroutines in the
differentiated modules do not have access to all variables of the
original module. Therefore the differentiated module must contain
its own copy of all the original module's variables, types, and procedures.
This is a change in TAPENADE's differentiation model: the differentiated code
cannot just call or use parts of the original code; it must contain its
own copies of those. In other words, the differentiated code need not
be linked with the original.

The interface mechanism of FORTRAN95 is a way to implement overloaded
procedures. This is static overloading, which is resolved at compile time.
Therefore we had to extend TAPENADE type-checking phase to completely
solve the calls to interfaced procedures. Conversely, TAPENADE is now able
to generate interfaces on the differentiated procedures, so that the general
structure of the code is preserved.

The array notation of FORTRAN95 is used systematically in OPA.
At the same time, differentiation requires that many calls to
intrinsic functions be split to propagate the derivatives.
When these functions are used on arrays ("elemental" intrinsics)
TAPENADE must generate a code which is far from trivial.
For instance the single statement from OPA:
\begin{verbatim}
zws(:,:,:) = SQRT(ABS(psal(:,:,:)))
\end{verbatim}
generates in the adjoint mode
\begin{verbatim}
abs1 = ABS(psal(:,:,:))
mask = (psal(:,:,:) .GT. 0.0)
...
WHERE (abs1 .EQ. 0.0) 
  abs1b = 0.0
ELSEWHERE
  abs1b = zwsb(:, :, :)/(2.0*SQRT(abs1))
END WHERE
WHERE (.NOT.mask(:, :, :)) 
  psalb(:, :, :) = psalb(:, :, :) - abs1b
ELSEWHERE
  psalb(:, :, :) = psalb(:, :, :) + abs1b
END WHERE
\end{verbatim}
Without going in too much detail into the adjoint differentiation
model, we observe that the test that is needed to protect the
differentiated code against the non-differentiability of {\tt SQRT} at 0,
as well as the test that controls the differentiation of {\tt ABS},
have been turned into {\tt WHERE} constructs to keep
the runtime benefits of array notation. Some temporary variables
are introduced automatically to store
control-flow decisions (e.g. {\tt abs1} and {\tt mask}),
although TAPENADE still doesn't do this in an optimal way on the example.

OPA uses pointers and dynamic memory allocation
(calls to {\tt ALLOCATE} and {\tt DEALLOCATE}). This is an application for
the pointer analysis now available in TAPENADE, finding
whether a variable has a derivative,
even when this variable is accessed through
a pointer. Unfortunately, dynamic allocation is handled partly, i.e. only
in the tangent mode of TAPENADE. In the adjoint mode, we have no general
strategy for memory allocation and TAPENADE sometimes cannot produce a working code.
We understand that the adjoint of an allocate should be a {\tt DEALLOCATE},
and vice-versa, but some changes must be made by hand on the
differentiated code to make it work.

\subsection{Checkpointing and hidden variables}

OPA reads and writes several data files, not only during the pre- and post-processing
stages, but also during the computational kernel itself.
Source terms such as the wind stress are being read at intermediate time steps.
Also, some modules and procedures define private {\tt SAVE} variables, whose
value is preserved but cannot be accessed from outside.
Although unrelated, these two points are just examples
of a common problem: they can make a procedure ``non reentrant".

If a called procedure modifies an internal {\tt SAVE} variable, it becomes impossible
from the outside calling context to call the procedure a second time with an identical
result. Similarly if the called procedure reads from a previously opened file, and
just moves the read pointer further in the file, then it becomes impossible
to call the procedure twice and obtain the same values read.

Non reentrant procedures are a problem for the checkpointing strategy of
the adjoint mode. We saw in section 3 that checkpointing relies on calling
the checkpointed piece twice, in such a way that the second call is
equivalent to the first. To this end, a sufficient subset of the
execution context, the snapshot, must be saved and restored.
Hidden variables like an internal {\tt SAVE} variable or the
read pointer inside an opened file cannot be saved nor restored
in general. When checkpointing would require hidden variables to be
put in the snapshot, then checkpointing should be forbidden.

Similarly, when a procedure only allocates some memory, the
allocation must not be done twice. If this procedure is checkpointed,
then one must deallocate the memory when restoring the snapshot
before the duplicate call.
TAPENADE is not yet able to do this automatically.

TAPENADE has some functionalities to cope with this hidden variables problem,
but in all cases interaction with the user is necessary.
First, TAPENADE issues a warning message when a subroutine
cannot be checkpointed because of a private {\tt SAVE} variable.
The message is issued only when this variable would be
part of the snapshot for this procedure.
When this happened for OPA, we just turned by hand the variables
in question into public global variables in the original code.
In principle this could also be done automatically.
However there are only a handful such variables,
thus developing this is not our priority.

When a subroutine is not reentrant because of I/O file pointers
or because of isolated memory allocation or deallocation, then
TAPENADE lets the user label the subroutine so that it must not
be checkpointed. For OPA, we took another strategy: we modified
the main I/O subroutines so that they always first make sure that
the file is opened and then only use direct read into the file
without using a read I/O pointer. Thus all I/O subroutines are reentrant.

\subsection{Binomial Checkpointing}

Automatic Differentiation of OPA is one of the most ambitious
applications of TAPENADE so far. It means building the adjoint
of a piece of code that performs an unsteady nonlinear simulation
over a very large number of time steps.
Each time step computes a new state whose size ranges in the hundreds
of megabytes. In adjoint mode if no checkpointing was applied, which means that
all intermediate values were to be stored on a stack, we could
execute only a handful of time steps before we run out of memory
even on our largest workstation.
Checkpointing is compulsory to compute the adjoint over
several thousands of time steps, which is our goal.

We saw in section 3 that TAPENADE applies checkpointing at
the level of subroutine calls, i.e. each call is checkpointed.
This easy strategy is often far from optimal.
On one hand several calls are better not checkpointed, and
TAPENADE now offers the option to mark selected calls for {\em not}
checkpointing.
On the other hand, checkpointing should be applied at other locations.
For example at the top level of
the simulation program is a loop over many time steps.
We definitely need an efficient checkpointing scheme applied
at this level of time iterations.

\begin{figure}[htbp]
\begin{center}
  \includegraphics[width=\linewidth]{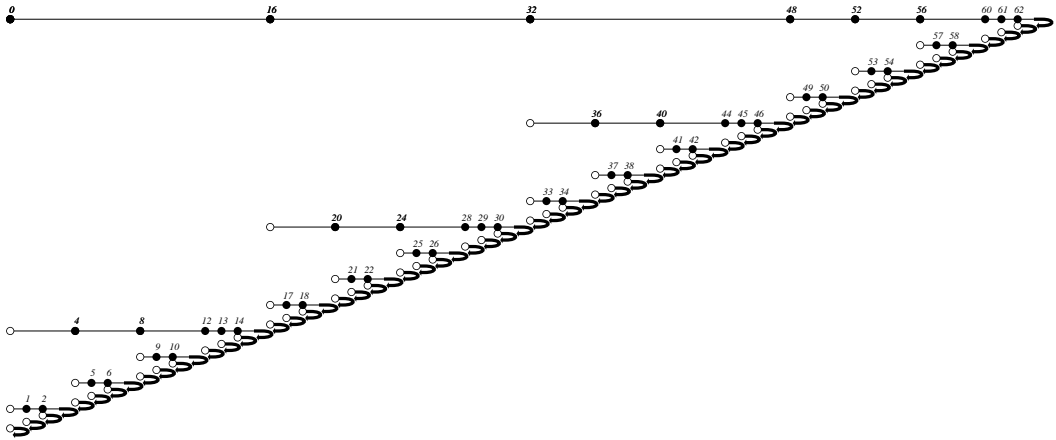}
  \caption{Three-levels checkpointing with 64 time steps and 9 snapshots.
Forward computations go right, adjoint computations go left.
Black circles represent writing/taking a snapshot, white circles
represent reading an available snapshot.}
  \label{figCkpMultiLevel}
\end{center}
\end{figure}

One classical solution used by TAF on the MIT GCM code~\cite{MIT-gcm},
is called  multi level recursive checkpointing.
Basically, it splits the complete time interval into
a small number of equidistant intervals, then apply the same
strategy to each of the sub-intervals.
For instance 64 time steps can be split into 4 large
intervals of 4 small intervals of 4 time steps,
as sketched on figure~\ref{figCkpMultiLevel}.
This consumes a maximum of 9 simultaneous snapshots,
and the average number of duplicate executions for a time step is $2.25$.
In a more realistic situation, 1000 time steps can be split into
10 large intervals of 10 small intervals of 10 time steps, and
one can figure out that
this consumes a maximum of 27 simultaneous snapshots,
and the average number of duplicate executions for a time step is $2.7$.

However, it was shown in~\cite{Walther-Greiwank}
that this strategy is not optimal.
Under the reasonable assumptions that all time steps cost the same
run time, and that the snapshot needed to run
again from time step $n$ to $n+1$ is the same as
to run from step $n$ to any later step $n+x$,
Griewank and Walther have characterized the optimal
distribution of nested checkpoints, which follows a binomial
law. With this optimal strategy, both spatial and temporal complexity
of the adjoint code grow logarithmically with respect to
the number of time steps of the original simulation.
In other words, both the slowdown factor which grows like the
number of times each time step is executed, and the memory
which grows like the number of simultaneous snapshots,
grow logarithmically with the total number of time steps.

\begin{figure}[htbp]
\begin{center}
  \includegraphics[width=\linewidth]{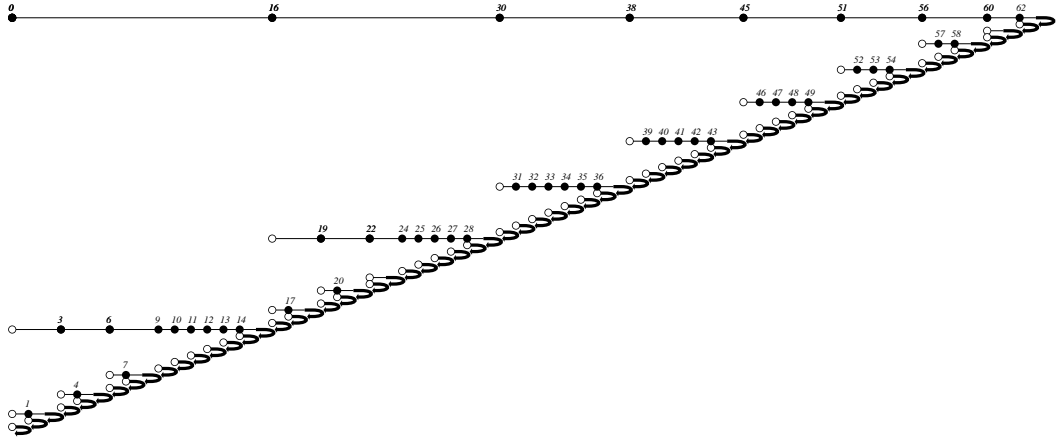}
  \caption{Optimal binomial checkpointing with 64 time steps and 9 snapshots}
  \label{figCkpBinomial}
\end{center}
\end{figure}

In real applications, run-time and memory space do not
behave symmetrically. One can always wait a little longer for the result,
whereas the memory space is bounded. Therefore the
maximum number of snapshots $d$ that can be stored simultaneously
is fixed.
Then~\cite{Griewank-1992} shows that the optimal strategy
gives a slowdown factor that grows only like the
$d^{th}$ root of the total number of time steps,
which is still very good.
Figure~\ref{figCkpBinomial} shows the optimal checkpointing strategy
for the same problem as figure~\ref{figCkpMultiLevel} i.e.
64 time steps with memory for 9 snapshots. The average
number of duplicate executions for a time step is only $2$.
For the more realistic situation (1000 time steps and memory for 27
snapshots) the average number of duplicate executions is only $2.57$.

\begin{figure}[htbp]
\begin{center}
  \includegraphics[width=6cm,height=14cm, angle=-90]{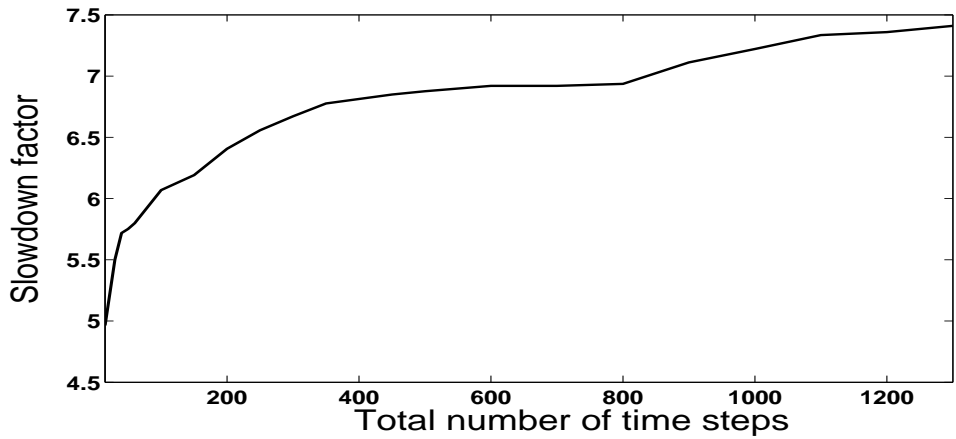}
  \caption{Optimal binomial checkpointing with 15 snapshots: slowdown factor as a function of the total length of the
  initial simulation.
  The slowdown factor is the run-time ratio of the adjoint code compared to the original code.
  }
  \label{FigureTreeverse}
\end{center}
\end{figure}

We implemented this optimal strategy in the adjoint code of OPA.
We made our first experiments by hand modification of the
adjoint code produced by TAPENADE. Still, TAPENADE produced automatically
the procedures that store and retrieve the snapshot, and therefore
the hand modification was benign:
given the number of time steps, a general
procedure\footnote{A FORTRAN95 implementation of this scheduling procedure can be found in
www.inria-sop/tropics/ftp/Hicham\_Tber/}
schedules the optimal sequence of actions
(store snapshot, retrieve snapshot, run time step, run adjoint time step)
to differentiate the complete simulation.
Further versions of TAPENADE will fully automate this process.
Figure~\ref{FigureTreeverse} shows the performances on OPA.
They are in good agreement with the theory.
Notice in particular the two small inflection points
on the curve around 150 iterations and 800 iterations.
Going back to the optimality proof in~\cite{Griewank-1992},
we see that the optimal strategy is particularly
efficient when the number of time steps is exactly
$$\eta(d,t) = \frac{(d+t)!}{d! t!}$$
where $d$ is the number of snapshots and $t$ is the
number of duplicate executions allowed per time step.
For our target machine $d=15$ and we find $\eta(15,2)=136$ and $\eta(15,3)=816$,
which corresponds to the inflection points of figure~\ref{FigureTreeverse}.

For the previous version OPA 8, the adjoint was written by hand.
Nevertheless, even a hand-written adjoint must implement
strategies to retrieve intermediate states in reverse order that is,
something very close to checkpointing.
Looking at this hand-written adjoint,
we first observe that the checkpointing strategy is
neither multi level nor optimal binomial. It is more like a single
level strategy, with one snapshot stored every fixed number of time steps.
During the reverse sweep, states between two stored snapshots
are rebuilt approximately using linear interpolation.
The advantage is that few time steps are evaluated twice,
and therefore the slowdown factor remains well below 4.
We can see at least two drawbacks. First, this hand manipulation
requires deep knowledge of the original program and of
the underlying equations. This method does not blend easily with
Automatic Differentiation. It is not yet automated
in any AD tool and therefore tedious and error-prone code
manipulations would still be necessary.
Second, this introduces approximation errors into the computed derivatives,
whose mathematical behavior is unclear. The gradient obtained in the end
is used in complex optimizations or data-assimilation loops, and small
errors may result in poor convergence.
In any case, for very large numbers of time steps, we believe
a trade-off between exact binomial checkpointing and approximate
interpolation is worth experimenting. Interpolation is probably
good enough for many variables that vary very slowly,
and which could be designated by the
end-user, and only the other variables would need to be stored.

\subsection{Iterative linear solver}

The OPA model solves an elliptic equation at the end of each
time step, using an iterative method
that generates a sequence of approximations of the exact solution.
The mechanical application of AD on this kind of methods gives a
sequence of derivatives of the approximate solutions with the
same number of iterations as the original solver.
The reason is that AD keeps the flow of control of the original
program in the differentiated program. In particular the convergence tests
are still based only on the non-differentiated variables.
Naturally, one may
ask whether and how AD-produced derivatives are reasonable approximations
to the desired derivative of the exact solution.
The issues of derivative convergence for iterative
solvers in relation to AD are discussed in \cite{Gilbert,DerivativeConv,Christianson}. 

\begin{figure}[htbp]
\begin{center}
  \includegraphics[width=\linewidth]{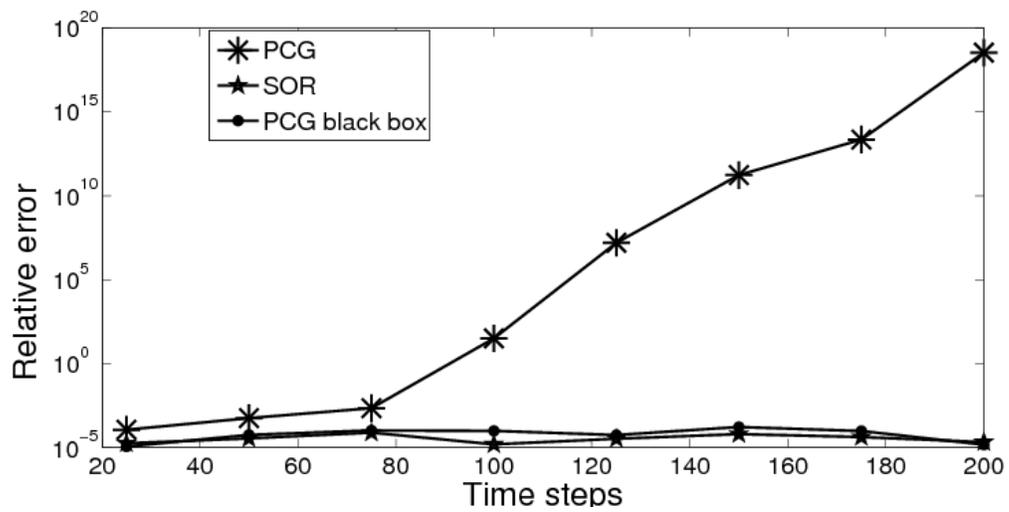}
  \caption{evolution of the relative error between tangent
derivative and divided differences,
for the three strategies: SOR and straightforward AD, PCG and straightforward AD,
PCG with the black box strategy}
  \label{FigSorVsPcg}
\end{center}
\end{figure}

OPA provides two alternative algorithms to solve the elliptic equation:
``PCG'' for Preconditioned Conjugate Gradient, and ``SOR'' for Successive Over-Relaxation method.
Both algorithms give correct results
for the original code, but PCG is generally preferred
thanks to its efficiency and vectorization properties.
However, the AD-differentiated code gives different results using the two algorithms.
Figure~\ref{FigSorVsPcg} compares the
AD-derivatives with approximate derivatives obtained by divided differences.
We see that the derivatives obtained with the SOR algorithm
remain correct when the number of time steps increases.
On the contrary, the derivatives obtained with the PCG algorithm
become completely wrong after 80 time steps.
Notice that this occurs in tangent mode as well as in adjoint mode:
 the  derivatives obtained with PCG, although wrong, remain identical in tangent and adjoint.
Our explanation is that each iteration
of PCG involves the computation of scalar products of variables that
depend on the state vector, thus making the numerical
algorithm nonlinear even though the elliptic equation is linear.
In~\cite{Gilbert}, Gilbert has shown that the application of AD to a fixed point
iteration gives a derivative fixed point iteration that converges
R-linearly to the desired derivative in particular in the case of
a large contractive iterate or secant updating.
Unfortunately this is not the case for quasi-Newton iterative solvers such as PCG,
for which there is no similar convergence result to our knowledge.

To solve this problem for the tangent-differentiated OPA we exploit the linearity of the elliptic
system, and for the adjoint-differentiated OPA we exploit the self adjointness property
of the elliptic operator \cite{WeaverEtAl}. We can thus use the original PCG routine
itself to solve for the differentiated linear systems.
Practically, we do this using the so-called
``black-box'' feature provided in TAPENADE.
Figure~\ref{FigSorVsPcg} shows that (here for the tangent mode)
the PCG gives the same accuracy as the SOR solver.

In another experiment, we tried to use straightforward AD with the PCG solver,
but this time fixing the number of PCG iterations to some very high value.
We observed that the derivatives become coherent again with divided differences.
This could be another way to solve our problem, but it is certainly expensive
and the choice of the ``high'' iteration number is delicate.
This problem definitely deserves further study, and confirms the
general recommendation not to differentiate solvers of a nonlinear kind,
and use a black-box strategy instead.

\section{Validation Experiments}

\subsection{Correctness test}

The classical way to check for correctness of the automatically generated tangent
and adjoint codes is as follows:
\begin{enumerate}
\item Choose an arbitrary input $X$ and and arbitrary direction $\dot{X}$.
Compute the Divided Difference
$$ D\!D = \dfrac{F(X+\varepsilon\dot{X})-F(X)}{\varepsilon} $$
for a good enough small $\varepsilon$.
\item Using the tangent differentiated program, compute
$\dot{Y} = F'(X) \times \dot{X}$.
\item Using the adjoint differentiated program, compute
$\overline{X} =  F'^{*}(X) \times \dot{Y}$
\end{enumerate}
and finally check that $(D\!D \cdot D\!D) = (\dot{Y} \cdot \dot{Y}) = (\overline{X} \cdot \dot{X}) $.
We performed this test for the complete global ORCA-2 simulation on 1000 time steps and its derivative codes.
The results are shown in table~\ref{TableDotProductTest}. The values match,
\begin{table}[htbp]
\caption{\label{TableDotProductTest}Dot product test for 1000 time steps}
\begin{centering}\begin{tabular}{|c|c|}
\hline 
$(D\!D \cdot D\!D)$ ($\varepsilon=10^{-7}$)&
 4.405352760987440e+08\tabularnewline
\hline 
$ (\dot{Y} \cdot \dot{Y}) $ &
4.405346876439977e+08\tabularnewline
\hline 
$ (\overline{X} \cdot \dot{X}) $ &
4.405346876439867e+08\tabularnewline
\hline
\end{tabular}\par\end{centering}
\end{table}
and $(\dot{Y} \cdot \dot{Y})$ and $(\overline{X} \cdot \dot{X})$ match very well, up to the last few digits,
which shows that the tangent and adjoint codes really compute the same derivatives,
only in a different computation order as shown by equations (\ref{eqtgtmode}) and (\ref{eqadjmode}).
The values of $(D\!D \cdot D\!D)$ and $(\dot{Y} \cdot \dot{Y})$ don't match so well,
because of the weakness of the Divided Differences approximation.
\begin{figure}[htbp]
\begin{center}
\begin{centering}\includegraphics[width=6cm,height=10cm, angle=-90]{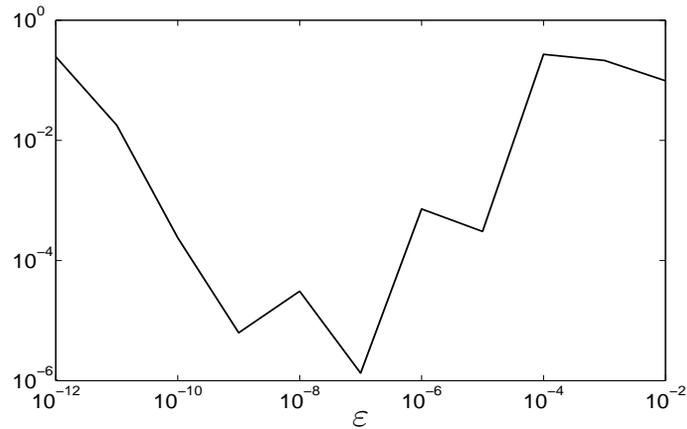}\par\end{centering}
\caption{\label{FigRelErrVsEps}Relative error of Divided Differences with respect to AD-generated derivatives,
computed for various values of of the step size $\varepsilon$}
\end{center}
\end{figure}
Figure~\ref{FigRelErrVsEps} shows this weakness:
For a small value of $\varepsilon,$ the dominant error is due to machine
accuracy. For a large value of $\varepsilon,$ the dominant error is
due to the second derivatives of $F$.
The best $\varepsilon$ minimizes both errors, but cannot eliminate them completely.

\subsection{Sensitivity analysis on a long simulation}
\begin{figure}[htbp]
\begin{center}
\begin{centering}\includegraphics[width=12cm,height=8cm]{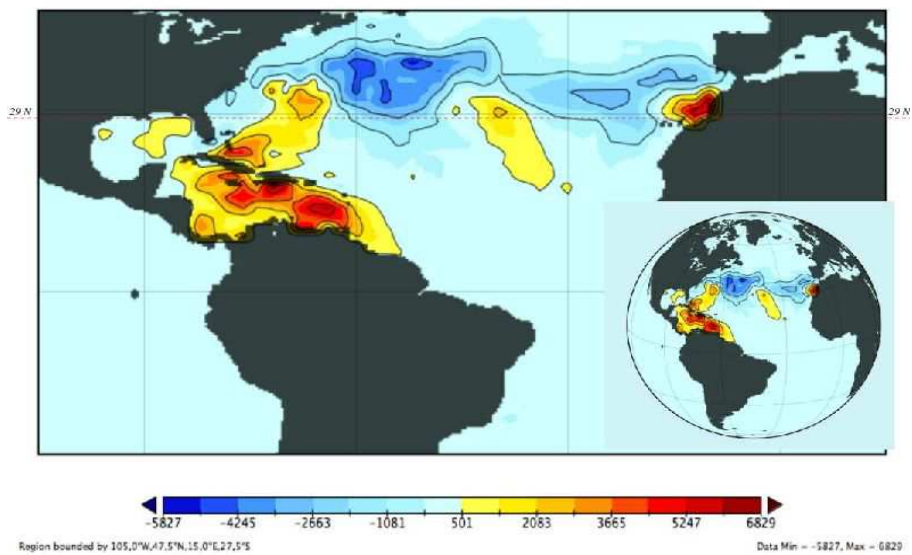}\par\end{centering}
\caption{\label{FigSensitivity}Sensitivity map of the North Atlantic
heat transport at 29°N (dotted line), with respect to changes in the initial surface temperature }
\end{center}
\end{figure}

 One of the main application of adjoint models is the sensitivity
analysis i.e. the study of how model output varies with changes in
model inputs. Using direct or statistical methods would require
many integration of the non linear model while one adjoint model
integration is enough to compute this sensitivity.
As an example, figure \ref{FigSensitivity} shows the output map of
the sensitivity of the North Atlantic meridional heat flux at 29°N
to changes in the initial sea surface temperature ($SST_{t_0}$)
over one year integration period, starting January 1, 1998.
This is done by computing the gradient respect to $SST_{t_0}$ of 
$$J = \int_{t_0}^{t_N}\iint_\Omega T.v\ dxdzdt$$ 
where $\Omega$ is the zonal cross section at  29°N in the North
Atlantic, $T$ is the temperature and $v$ is the meridional current velocity.

Contours in figure \ref{FigSensitivity} show where variation of
initial SST would effect the most upon heat transport at 29°N.
It shows large scale patterns mainly located north of the 29°N
parallel and in the Caribbean sea with a strong spot off Morocco.
These results are consistent with those obtained by
Marotzke et al. (\cite{Marotzke1999Cot})

This map was computed by the TAPENADE-generated adjoint of OPA on
the global ORCA2 grid, over 5475 time steps (1 year).
This experiment was done with the SOR algorithm as the iterative
linear solver.
The TAPENADE-generated adjoint computed this sensitivity map
in a time that is only 8.03 times that of the original simulation.

\subsection{Data Assimilation}

For further validation of the automatically generated derivatives,
we carried out a data assimilation experiment.
This was done in a so-called twin experiment framework
whereby the direct model trajectory is used to generate
synthetic observations.
The initial sea surface temperature is perturbed by a
white noise and it has to be recovered using variational
data assimilation techniques. Synthetic observation are
given by the sea surface height (SSH) and the sea surface
salinity (SSS) generated from the model's original outputs
starting from the unperturbed SST.

The cost function to be minimised is 
\begin{equation}
J(SST({t_0}))=\int_{t_0}^{t_N}\parallel SSH(t)-SSH^o(t)\parallel^2+\parallel SSS(t)-SSS^o(t)\parallel^2 dt
\end{equation}
Where the superscript $^o$ stands for synthetic observation and $SSH(t)$ and $SSS(t)$ are model output.

\begin{figure}[htbp]
\begin{center}
\begin{centering}\includegraphics[width=10cm,height=6cm]{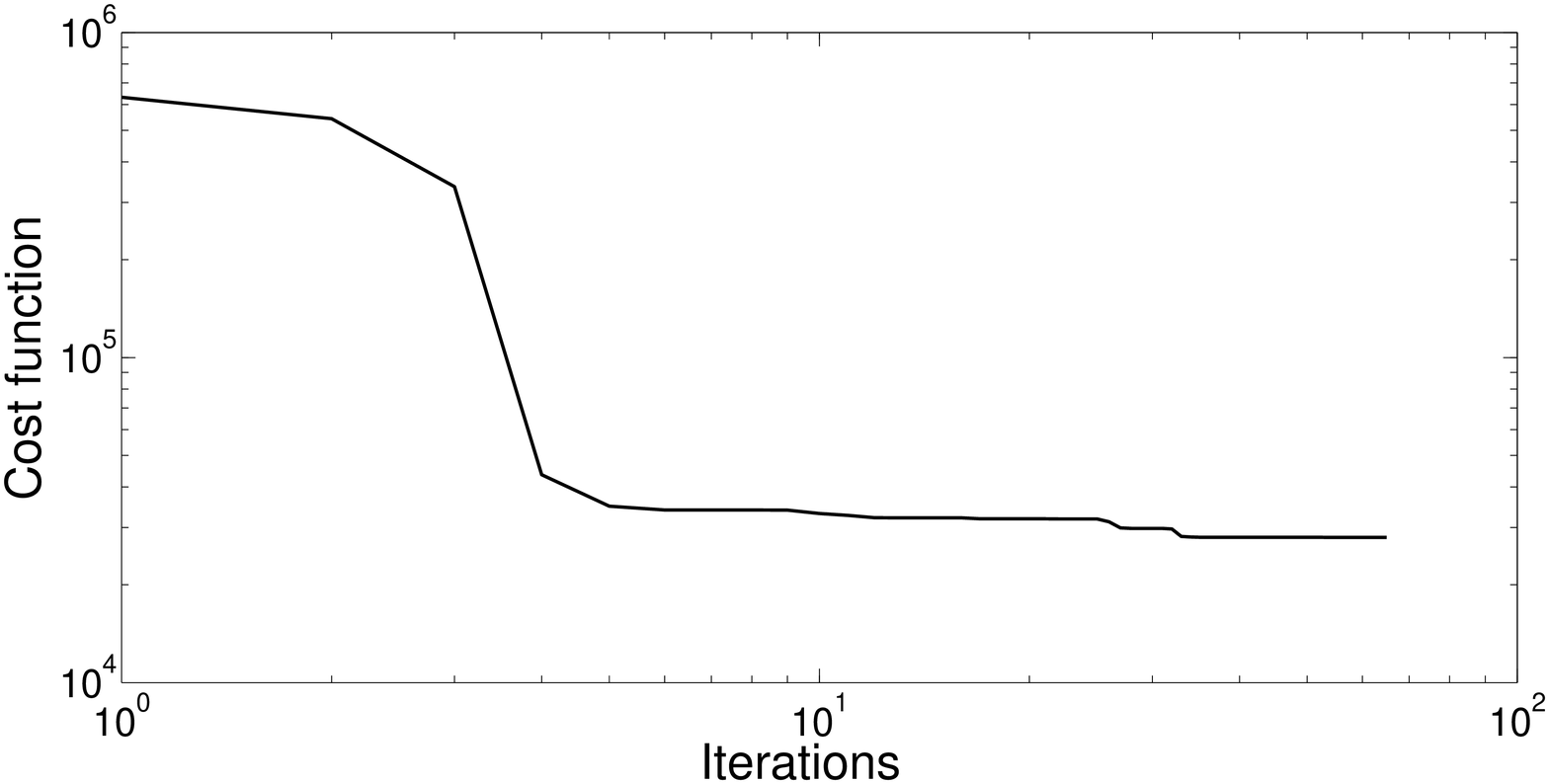}\par\end{centering}
\caption{\label{CostFunctionVsIterations}Twin experiment: Convergence of
the cost function}
\end{center}
\end{figure}

For computing cost issues, only the Antarctic zoom of ORCA2
is considered, the minimisation is done an iterative gradient
search algorithm where the gradient of $J$ is cumputed using
adjoint techniques. Figure \ref{CostFunctionVsIterations}
illustrates the performance of the optimization loop for
an integration period of 1 month i.e. 450 time steps.
The cost function decreases by two orders of magnitude.

\begin{figure}[htbp]
\begin{center}
\begin{centering}\includegraphics[width=14cm,height=10cm]{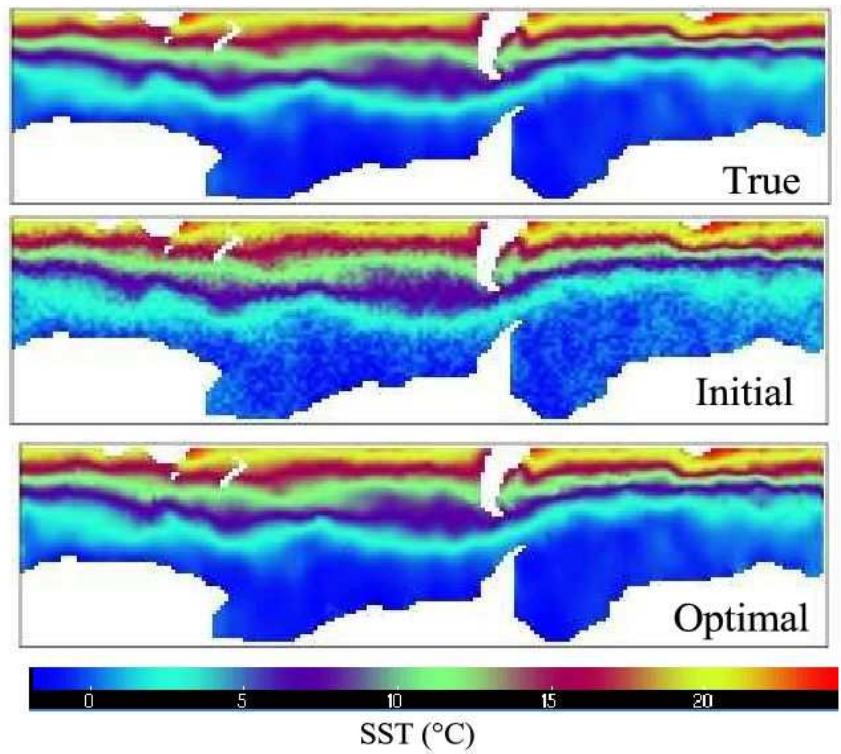}\par\end{centering}
\caption{\label{TrueInitialOptimal}Twin experiment: True field (top), Initial perturbed field (middle) and
identified optimal sea surface temperatures (bottom)}
\end{center}
\end{figure}

Figure~\ref{TrueInitialOptimal} indicates that the true solution (top panel) is
recovered with a good approximation (bottom panel) from the randomly perturbed one(middle panel), showing the quality of the derivatives
obtained.

\section{Conclusion and Outlook}

The effort to build the tangent and adjoint codes for the
previous version 8 of the OPA ocean General Circulation Model
has cost several months development from an experienced researcher.
For the new version OPA 9 written in FORTRAN95, the use of the AD tool
TAPENADE significantly reduces this effort.
Our first numerical applications show the quality of the derivatives
obtained.
This works validates the choice of AD as the strategy to obtain
the tangent and adjoint for OPA 9, and for the versions to come.

At the same time, OPA is the largest FORTRAN95 application
differentiated with TAPENADE. This work has pointed at a number of
limitations of TAPENADE that have been lifted.
Other limitations remain, such as the non-reentrant procedures,
which need to be addressed in future work.
Successful differentiation of OPA definitely increases our confidence
in TAPENADE.

This works is also an additional illustration of the superiority
of the binomial checkpointing strategy, compared to multi level
checkpointing. By the standards of other application fields,
e.g. CFD, a slowdown of the adjoint code of only 7 for
a nonsteady simulation on 1000 time steps would be considered very good.
By the standards of weather simulation or ocean modeling however,
scientists expect yet faster adjoints, at the cost of a
radical approximation. Even if we consider that these approximations
change the mathematical nature of the optimization process,
we understand they are necessary and we shall study how they can be
proposed as an option by the AD tool.

This work has underlined several directions for further research
in AD and AD tools. Some of them are already being studied by researchers
in our groups. Considering the application language, two constructs
need to be differentiated better:
\begin{itemize}
\item The next experiment to be made very soon is to apply
TAPENADE to the parallelized version of OPA. This is necessary before
the generated tangent and adjoint codes can be used in production
context.
\item The OPA source makes extensive use of the preprocessor directives
such as {\tt \#IFDEF}. TAPENADE does not deal with these directives
because they do not respect the syntactic structure of a code.
Handling these directives in the AD tool is in our opinion hopeless.
What might be done though, is to generate differentiated codes
for each possible preprocessed code, and devise a tool to
put the directives back into the differentiated codes.
This is made easier if the differentiated code closely
follows the structure of the original, as is the case with TAPENADE.
\end{itemize}
Considering specifically adjoint differentiation, we hope to
obtain more efficient code through a more systematic exploitation of
self-adjointness, e.g. of the elliptic operator. We also hope to
optimize the checkpointing strategy. In its present version,
TAPENADE applies checkpointing to each procedure call.
Using profiling information, we believe we can detect several
procedure calls for which checkpointing is useless or counter-productive.
TAPENADE is already able to use this information to produce a better adjoint.


\bibliographystyle{plain}
\bibliography{InriaReport}

\end{document}